\documentclass[10pt,sigconf,letterpaper,nonacm]{acmart}

\usepackage{stfloats}
\usepackage{subcaption}
\usepackage{soul}
\usepackage{multirow}
\usepackage{xspace}
\setcopyright{none}
\usepackage{xcolor}
\usepackage{datetime2}
\DTMsetdatestyle{iso}   %
\makeatletter
  \newcommand\EatSpacesHack{\@bsphack\@esphack}
\makeatother

\newcommand{\WIMIA}{\url{WhatIsMyIPAddress.com}\xspace}
\newcommand{\ZeroAst}{\makebox[0pt][l]{*}}

\def\Snospace~{\S{}} %
\acmConference[IMC '26]{Make sure to enter the correct
  conference title from your rights confirmation email}{Oct 28--31,
  2025}{Madison, WI}

\title{Smoothing Rough Edges of IPv6 in VPNs}

\author{Yejin Cho}
\affiliation{%
  \institution{University of Southern California}
  \city{Los Angeles}
  \country{USA}
}
\email{yejincho@usc.edu}

\author{John Heidemann}
\affiliation{%
  \institution{Information Sciences Institute}
  \institution{University of Southern California}
  \city{Los Angeles}
  \country{USA}
}
\email{johnh@isi.edu}

\copyrightyear{2025}
\acmYear{2025}
\setcopyright{rightsretained}
\acmConference[IMC '25]{Proceedings of the 2025 ACM Internet Measurement
Conference}{October 28--31, 2025}{Madison, WI, USA}
\acmBooktitle{Proceedings of the 2025 ACM Internet Measurement Conference
(IMC '25), October 28--31, 2025, Madison, WI, USA}
\acmDOI{10.1145/3730567.3768612}
\acmISBN{979-8-4007-1860-1/2025/10}

\begin{abstract}
How do commercial VPNs interact with IPv6?
We show two ``rough edges'' in how commercial VPNs handle IPv6.
First, we show that many \emph{IPv4-only VPNs leak IPv6 traffic to the ISP}.
Individual use VPNs in part to conceal their local IP addresses,
  so such leaks reduce user privacy.
While prior work has studied VPNs in testbeds,
  we use a new dataset of 129k VPN-using daily visitors to \WIMIA
  that quantifies these leaks and show 12
  VPNs previously considered safe still leak for at least 5\% of their users.
We show native IPv6 addresses leak most commonly
  in VPNs that claim only IPv4 support,
  with 5\% to 57\% of visitors of v4-only VPNs having their native IPv6 address exposed.
Second, we show that \emph{most dual-stack VPNs users
  actually select IPv4 instead of IPv6}.
We observe this problem in our visitor data,
  and we identify the root cause arises
  because when user's computer
  follows standard address-selection rules,
  VPN-assigned addresses are
  often de-preferenced.
Testing six VPNs on Android, we show that five consistently de-prioritize IPv6. 
Finally, 
  \emph{we suggest a solution to IPv6 de-preferencing}: we
  define a new IPv6 address range for VPNs
  that is not de-preferenced by address selection.
We prototype this solution on Linux. 
Our findings help identify and address rough edges
  in the addition of IPv6 support to VPNs.
\end{abstract}

\begin{document}

\maketitle

\section{Introduction}

Virtual private networks (VPNs) and IPv6 are important parts of today's Internet.
VPNs improve privacy and protect against censorship
  by encrypting data and hiding end-user IP addresses from eavesdropping.
IPv6 offers enough addresses to enable true peer-to-peer, end-to-end connectivity.
VPN providers increasing advertise native IPv6 support
  as a key feature~\cite{proton_supports_ipv6,hideme_supports_ipv6, mullvad_supports_ipv6}.

Our paper explores \textbf{how commercial VPNs actually interact with IPv6}.
We find two ``rough edges'' in practice:
  \emph{IPv4-only VPNs often leak IPv6 traffic}, failing to protect user privacy,
 \emph{VPNs advertise IPv6 support as feature but actually connects over IPv4}, even though IPv6 is available and working.
Wee explore these rough edges with a novel data source,
  \WIMIA.

Our first contribution is to quantify
  \textbf{how often VPNs leak IPv6 traffic}, a rough edge
  where VPNs fail to provide
  the privacy they promise users.
VPNs are designed to hide a user's local identity and IP addresses,
  rewriting them with VPN addresses when traffic leaves the VPN to its destination.
If an IPv6 address from the user's computer leaks on to the Internet,
  that provides a unique identifier that may allow a
  connection to the users real-world identity.

We use a new data source, \WIMIA,
  to look at what IPv4 and IPv6 addresses users \emph{actually} expose
  (\autoref{sec:ipv6_leak}).
We show that, in practice between 5 and 57\% of all IPv4-only VPN users
  expose a local IPv6 address.
While prior work has examined VPNs for IPv6 leaks
  based on testbed experiments~\cite{perta_glance_2015,ikram_analysis_2016,khan_empirical_2018,ramesh_vpnalyzer_2022},
  our use of real-world data allows us to see
  leaks that occur potentially due to user misconfiguration or OS bugs.

Our second contribution is to show that
  \textbf{most dual-stack VPNs users
  actually select IPv4 instead of IPv6},
  even when both are available from the user and the destination (\autoref{sec:ipv6_depref}).
We observe this problem in our \WIMIA data.
We show that the root cause is that the user's computer follows the standard address-selection rules when examining local network interfaces, while the VPN uses a ULA IPv6 address. 
The standard address-preferencing rules~\cite{rfc6724} will always 
select the private IPv4 address over this IPv6 ULA address, 
for reasons we explain in \autoref{sec:ipv6_depref_cause}.

Our final contribution is to show
  \textbf{solutions to these rough edges}.
We suggest that all VPNs need to support IPv6 so they can protect
  IPv6 addresses appropriately (\autoref{sec:leak_fix}).
We suggest a 
  a new IPv6 address class
  can solve IPv6 de-preferencing (\autoref{sec:ipv6_depref_solution}).
We demonstrate this change through an implementation in Linux.  

\textbf{Ethical Considerations and Data Availability}:
Our study was reviewed by
 % \ifisanon
  our university's IRB (number anonymized),
 % \else
 % \fi
  which classified it as Not Human Subjects Research (NHSR).
The client data we analyze consist of IP logs collected during ordinary website visits, in line with the site's privacy policy.
We discuss ethics in \autoref{sec:ethics}.
As part of our data use agreement, our \WIMIA data is not available,
  but we provide 
  data from our testbed
 % \ifisanon
  experiments~\cite{OurDataAnon}.
 % \else
 % \fi

\section{Background and Related Work}
	\label{sec:related_work}

\subsection{Background: Dual-Stack Connection}
	\label{sec:background}

Dual-stack operation,
  when both the source and destinations support both IPv4 and IPv6,
  is the primary transition strategy for IPv6 today.

With years of transition experience,
  connections today follow three steps
  to select a protocol with minimal user latency:
  (1) DNS resolution,
  (2) address selection, and
  (3) connection racing.
 These steps are implemented jointly by the application (typically a web
browser) 
and the operating system (networking libraries and the
kernel); 
we describe the behavior of popular browsers in
\autoref{sec:ipv6_depref_browser_details}.

Suppose, a dual-stacked user wants to visit dual-stack website. 
First, the browser issues DNS queries to obtain a set of destination
addresses (both A and AAAA records).
Second, for each destination address (A and AAAA), the host chooses an appropriate
source address and interface
  and \emph{prioritizes} the resulting (source, destination) pairs
  into a ranked list according to its address-selection policy~\cite{rfc6724}.
We describe this policy in detail in \autoref{sec:ipv6_depref_prioritzation_rules_explained}.
Finally, if there multiple protocol options remain,
  the application may use the Happy Eyeballs algorithm~\cite{schinazi_happy_2017}
  to select a quickly responding protocol.
Although browsers differ in their implementations of these steps,
  the results we report 
  are due to the RFC specifications
  and repeat across all implementations.

\subsection{VPN Privacy and IPv6 Leakage}
	\label{sec:related_work_leaks}

Several prior studies have evaluated IPv6 traffic leakage from VPN providers
  in 2015~\cite{perta_glance_2015},
  2016~\cite{ikram_analysis_2016},
  2018~\cite{khan_empirical_2018}, 
  and 2022~\cite{ramesh_vpnalyzer_2022}
  showing VPN evolution over time.
These studies show VPNs are improving
  and leaking less IPv6 traffic.
The first studies showed nearly all VPNs leaked IPv6
  (14 of 14 in 2015, and 56 of 67 in 2016),
  with far fewer in the last two studies
  (12 of 43 in 2018 and 5 of 80 in 2022).
However, all prior work evaluated VPNs as leak or not-leak;
  we instead consider the percentage of sessions we see that leak
  and show that several VPNs prior studies identified as ``safe'' do leak sometimes.

Each of these prior studies uses a similar methodology:
  they install and test selected VPNs
  by monitoring 
  device-level traffic for leaks.
They differ by adding more VPNs,
  or more operating systems.
  
This methodology suffers from combinatorial explosion
  with five major OS platforms and dozens of VPNs.
Our work instead uses data from a popular website to study
  actual VPN users, providing insight into many platforms
  (more than can be tested by hand in prior work).
Like their work, we test specific software to confirm our results
  and test our proposed changes.
We provide a complete list of comparisons of VPN providers in ~\autoref{sec: comparison_table}.

\subsection{Evaluation of Happy Eyeballs and IPv6 Preferencing}

Prior work has examined failures in Step~1 and Step~3 of \autoref{sec:background}.
Some studies focus on \emph{DNS behavior} (Step~1)---for example, resolver bias, delayed AAAA responses, or inconsistent dual-stack DNS handling~\cite{apnic_ipv6}.
Other studies focus on \emph{Happy~Eyeballs behavior} (Step~3), showing that faulty browser implementation can cause IPv6 de-preference even when IPv6 is otherwise functional~\cite{sattler_lazy_2025}.

In contrast, our work focuses on the \emph{address-selection stage} (Step~2)---the stage that precedes Happy~Eyeballs.
We show that local addressing configurations
  in commercial VPNs can shift RFC~6724 outcomes and cause systematic IPv6 de-preference, even when:
(i) DNS resolution returns valid AAAA records, and
(ii) Happy Eyeballs would correctly prefer IPv6 if given a properly ordered candidate list.

\section{Data Source}
	\label{sec:data-source}
        
To get a sense of how VPNs interact with IPv6 in the real-world,
  we require a data source that samples millions of users
  of VPNs with both IPv4 and IPv6 address observed.
We use data collected by \WIMIA,
  a website millions of people use to identify their IP addresses.
This website responds to users evaluating 
  their IPv4 and IPv6 addresses, if any.
Although it reports computers and their IP addresses,
  it does not collect any information about the users of those computers,
  and we agree not to attempt to identify individuals in this data
  (consistent with our IRB).

When a computer accesses \WIMIA,
  it connects to the website using either IPv4 or IPv6
  based on some algorithm,
  often a version of Happy Eyeballs.
The website then
  identifies the other protocol
  by returning HTML with embedded resources
  that use only other IP protocol,
  allowing it to identify both addresses and log which protocol is preferred.

We use data from \WIMIA covering 30 days (5.2 million sessions,
  of which 4 million are unique after removal of repeats in each hour),
  from 2025-02-22 to 2025-03-24.
  
\textbf{Supplemental data:}
We augment \WIMIA data with
  VPN detection and classification from IPinfo~\cite{ipinfo_vpn},
  organizations from CAIDA's AS2Org~\cite{caida_as_orgs},
  and AS classifications from ASdb~\cite{ziv_asdb_2021}.

\section{How Often Do VPNs Leak IPv6 Traffic?}
	\label{sec:ipv6_leak}

VPNs are intended to protect user privacy and bypass censorship by encrypting traffic and tunneling through a VPN server.
However, some VPNs fail to tunnel IPv6 traffic, causing to 
exit through the user’s native network interface and expose the real IPv6 address.

\subsection{Methodology: Detecting Leaks}
	\label{sec:ipv6_leak_detection}

We define an \emph{IPv6 leak} 
  on a dual-stack device
  as when
  a user's IPv4 traffic uses the VPN,
  but their IPv6 does not.
Without the VPN for IPv6,
  the user does not get the privacy they expected:
  their non-VPN'ed IPv6 address is externally visible,
  and their traffic risks exposure because it is not encrypted by the VPN.

We \emph{observe} IPv6 leaks from paired IP logs our
  \WIMIA data (\autoref{sec:data-source}):
We detect a VPN when the IPv4 address is associated with a VPN service,
  and we identify a \emph{VPN leak} when the IPv6 address
  (i) does not belong to a VPN service, and
  (ii) originates from a different organization, and
  (iii) is classified as ISP AS.
We determine using 
  (i) IPinfo VPN classification~\cite{ipinfo_vpn},
  (ii) from CAIDA's AS2Org~\cite{caida_as_orgs},
  and (iii) from ASdb~\cite{ziv_asdb_2021}.

 % \iffalse
 % \fi

\begin{figure*}
    \begin{subfigure}[b]{\textwidth}
    \mbox{\includegraphics[width=0.95\linewidth,trim=0 30 0 0]{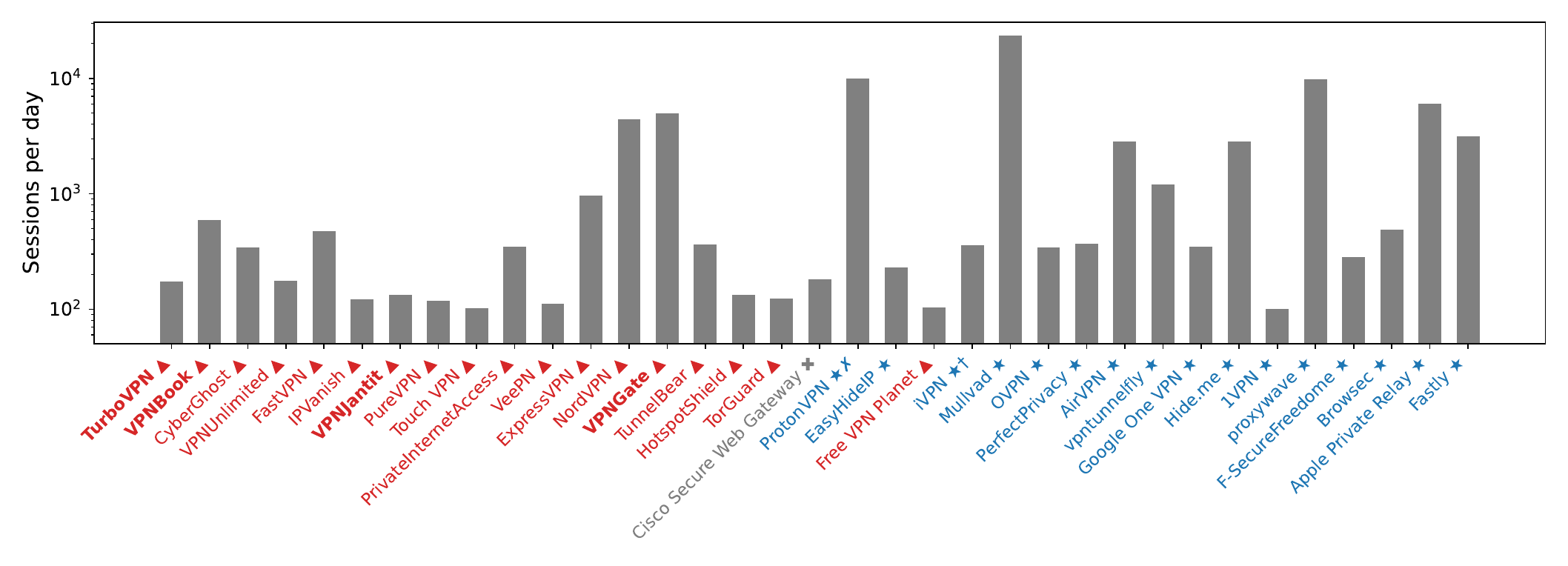}}
    \caption{Mean number of VPN sessions (log-scale) per day, for each VPN provider.}
        \label{fig:total_traffic_height_only}
    \end{subfigure}

    \begin{subfigure}[b]{\textwidth}
    \mbox{\includegraphics[width=0.95\linewidth,trim=0 30 0 0]{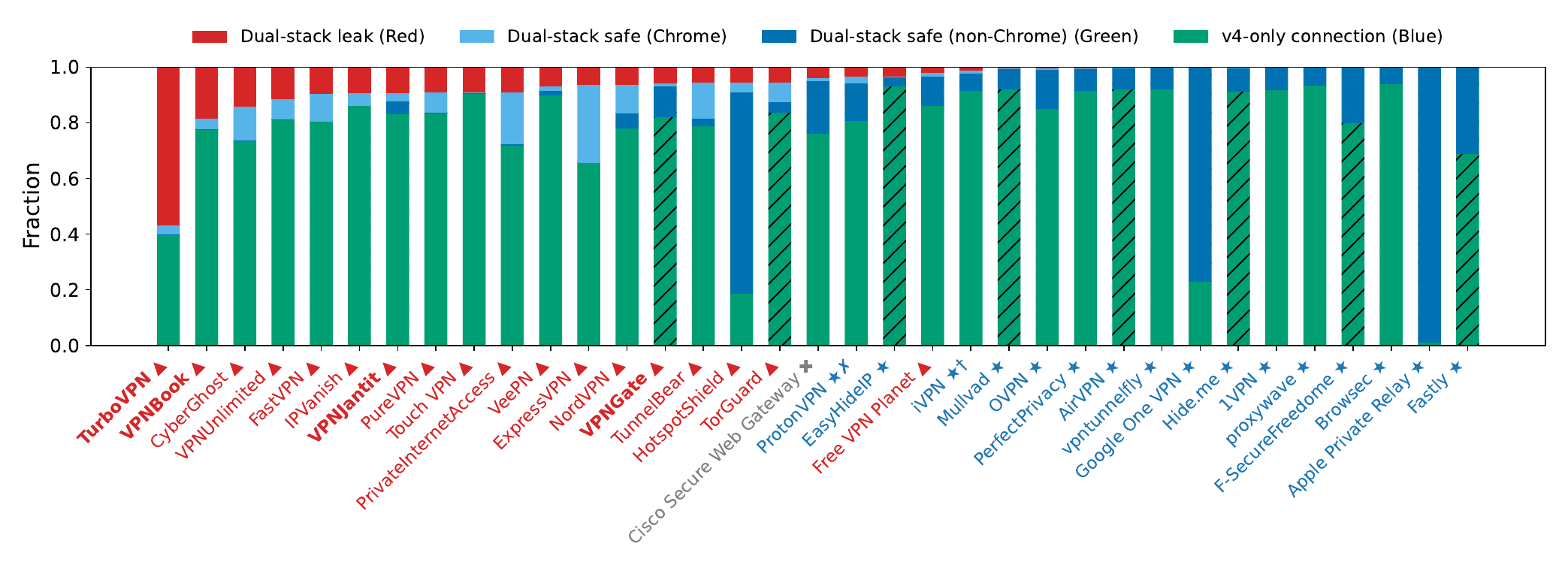}}
    \caption{Distribution of session types (v4-only, dual-stack safe non-Chrome and Chrome, and dual-stack leaking) for each VPN provider. }
    \label{fig:total_traffic_normalized}
    \end{subfigure}

    \begin{subfigure}[b]{\textwidth}
    \mbox{\includegraphics[width=0.95\linewidth,trim=0 30 0 0]{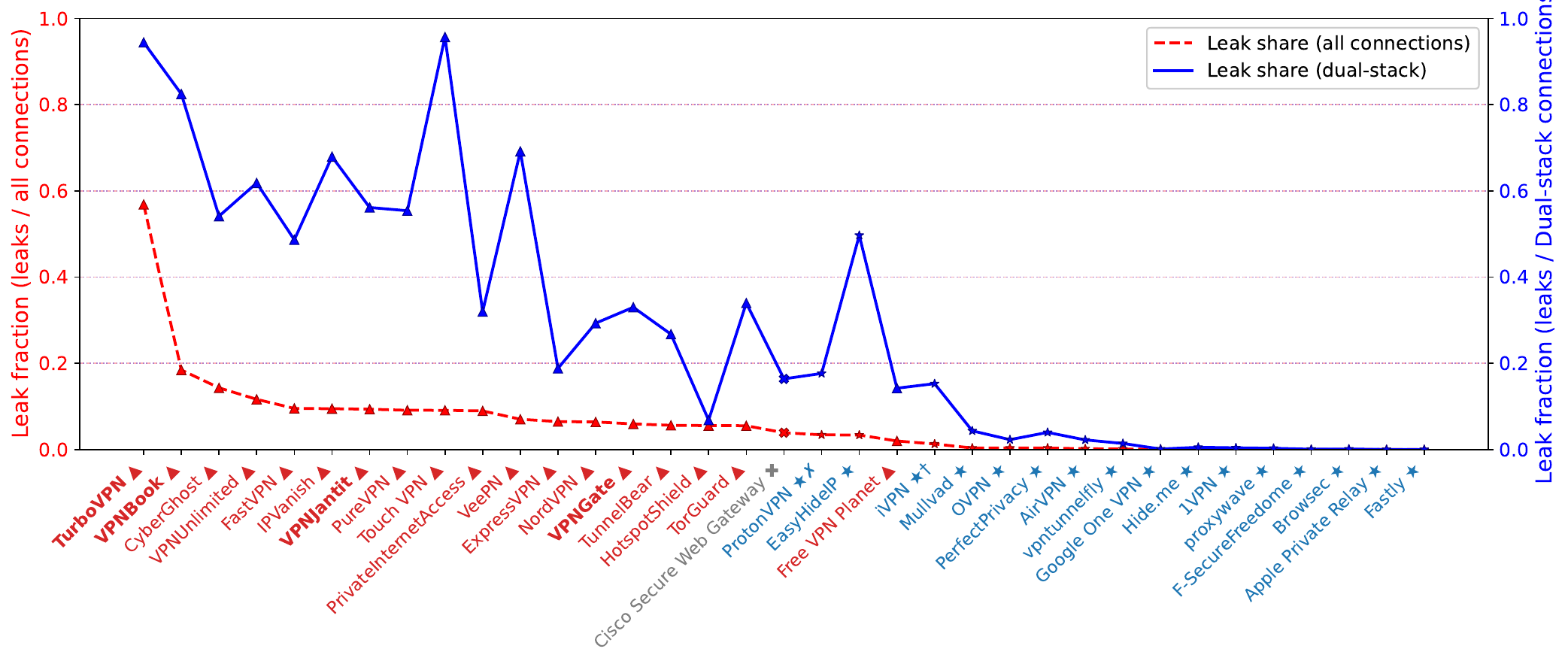}}
    \caption{Fraction of \WIMIA sessions with IPv6 leaks,
    shown relative to all sessions (the lower, red, dashed line)
    and dual-stack sessions (the upper, dark blue, solid line).}
    \label{fig:vpn_leaks}
    \end{subfigure}

\caption{Evaluation of \WIMIA data and IPv6 leaks.
    VPN providers are shown on the $x$-axis,
    sorted by their leak rate over all sessions (the dashed red in \autoref{fig:vpn_leaks}).
    VPN type: v6-supported (VPN name blue with stars),
  v4-only (red with triangles),
  and operator-policy-dependent (gray with an X).}
    \label{fig:ipv6_leaks}
\end{figure*}

\subsection{Basic Traffic and IPv6 Leaks Across All Users}

We first look at overall traffic and the rate of leaked 
IPv6 addresses.
Our \WIMIA dataset spans 30 days and 4M users after preprocessing (\autoref{sec:data-source}).

We identify VPN users by IPv4 addresses that come from a known VPN address,
  as determined by IPinfo~\cite{ipinfo_vpn}.
\autoref{fig:total_traffic_height_only} shows the mean number of sessions
  (log-scale)
  for each day, broken out by VPN.
We detect 123 VPNs in our data,
  and report only the 35 VPNs that have more than mean of 100 sessions per day after cleaning.
Here we sort VPNs by overall IPv6 leak rate,
  and label dual-stack (names are blue with stars),
  v4-only (red with triangles),
  and operator-policy-dependent (gray with an X).
  We manually determine IPv6 support based on the information provided on each VPN service’s webpage.

We then identify IPv6 leaks as described in \autoref{sec:ipv6_leak_detection}.
\autoref{fig:total_traffic_normalized} shows the distribution of connections by category.
The top red bar represents the fraction of VPN connections that leak IPv6.
The next, sky-blue bar corresponds to VPN connections that are safe because the IPv6 traffic is from Chrome’s prefetch server.
Below that, the dark-blue bar shows dual-stack connections where both IPv4 and IPv6 are correctly tunneled through the VPN, and are therefore safe.
Finally, the bottom light-green bar represents IPv4-only connections from VPN.

\textbf{Leak rate across all users of VPNs:}
In \autoref{fig:vpn_leaks}, the dashed red line
  shows the fraction of IPv6 leaks that occur across all visitors
  to \WIMIA.
We see that leaks range from about 0\% to 30\% based on VPN package.
IPv4-only VPNs leak more frequently
(mean 6.5\%, median 2.8\%)
than dual-stack VPNs (mean 2.9\%, median 1.8\%)
or unidentified services (mean 0.9\%, median 0.4\%)

Looking at all sessions shows that most IPv4-only VPN packages protect their users
  because they \emph{do} usually disable IPv6,
  but sometimes they don't.

\subsection{IPv6 Leaks from Dual-Stack Visitors}
We next focus on the subset of VPN users who are dual-stacked.
We find that, although we would normally expect all dual-stacked connections from IPv4-only VPNs to be a leak,
the actual leak rate varies.

The top solid blue line in \autoref{fig:vpn_leaks} in gives the
  fraction of VPN users, who are dual-stacked, (those we observe both v4 and v6 addresses)
  who leaks native IPv6 addresses, grouped by VPN operator.
We first consider the percentage of leaks
  relative to all dual stack users in the solid red line.
We observe that
  VPNs with IPv6 support (blue stars)
   show lower leak rates 
 than those that are IPv4-only (red triangles).
We are surprised that
  some IPv4-only providers sometimes do protect IPv6 traffic,
  showing leak rates %
  below 100\% (ranging from about 95\% down to 5\%).
We find that some self-identified IPv4-only providers still protect IPv6 traffic, with leak rates as low as 5\%. 
Hotspot Shield is a striking example: although it identifies as an IPv4-only VPN and its customer support claims it ``does not currently support IPv6,'' our measurements show that roughly 95\% of its dual-stack traffic is nevertheless protected.
We verified that all `safe' traffic originates from the same AS for both IPv4 and IPv6. %

Two special cases are relevant to our classification:
First, we sometimes observe partial IPv6 deployment in some VPNs clients.  
These cases do not satisfy our leak criteria (i to iii), because the IPv6 address does belong to the VPN's AS, 
  so they appear as unexpected protection, not a leak.  
Second, some traffic classified as safe, dual-stack sessions 
  is due by Chrome prefetching.
The Chrome browser speculatively loads resources 
  in advance to improve page load performance.
When it does so,
  the initial fetch of main page comes from Chrome's AS,
  but subsequent requests %
  follow through the user's VPN. 
Thus the prefetched IPv6 connection appears outside the VPN
  and would be misclassified as a leak.
To correct this case
  we treat traffic from prefixes assigned to Chrome Prefetch~\cite{chrome_prefetch_download} as safe.

\subsection{Testbed Validation and Recommendations}
    \label{sec:leak_testbed_validation}
    \label{sec:leak_fix}

To confirm our \WIMIA findings, we tested few free VPNs in a controlled testbed. 
We installed TurboVPN, VPNBook, VPNJantit, VPNGate, and TunnelBear (listed in decreasing order of IPv6 leakage from our analysis) and found that all but TunnelBear leaked IPv6 traffic. 
This validates that the IPv6 leakage behaviors observed in our dataset also occur in real-world VPN clients.
We list the experiment environment for each VPN in the \autoref{sec: test_vpn_specifications}.

Our recommendation to address IPv6 leaks is that VPNs
  should fully support IPv6,
  as is shown by the very few leaks we see in dual-stack VPNs (blue VPNs marked with a star in \autoref{fig:vpn_leaks}).

\subsection {Limitations in Studying IPv6 Leaks }

Although our data from \WIMIA captures a broad view of IPv6 traffic and potential
  leaks, it is not perfect.

First, \WIMIA is often used by users debugging networking problems.
We do not claim that it represents the general Internet population,
  but only that it represents a broad population of millions of users.

Second, some leaks may be because users explicitly
  modified their network configuration.
Thus even though the VPN disables IPv6,
  the user overrides this safe setting.
Although these cases are real IPv6 leaks,
  user override should not be attributed as a limitation fault of the VPN software.
Potentially such overrides may be more common in our data
  if \WIMIA attracts more technically savvy users.

Finally, our work assumes the supplement data
  we use from external sources is correct.
Misclassification in VPN classification data
  could bias which providers are included in our study.
  
In fact, in a discussion with one VPN company (anonymized),
  we learned that the database we used covered only about 1\% of their actual VPN exit IP addresses, which means our findings may not linearly reflect the number of VPN leaks in the real world.
In addition, the ASdb database was last updated in May 2021,
  while our work was done in 2025, so it may not fully reflect current ASes.

\section{Do VPNs De-preference IPv6?}
	\label{sec:ipv6_depref}

When users have both IPv4 and v6, they have an option which protocol to use.
Happy Eyeballs favors IPv6 (\autoref{sec:related_work}),
  but we show that in practice,
  5 out of 6 VPNs favor IPv4.
We show the cause of VPN de-preferencing
  and suggest how this problem can be fixed.

\subsection{Observing IPv6 De-Preferencing in the Wild}

Happy Eyeballs (\autoref{sec:related_work})
  is designed to favor IPv6 when given a choice of protocol.
We evaluated the protocol choice in \WIMIA.

\begin{figure*}
    \includegraphics[width=\linewidth, trim=8 0 0  0, clip]{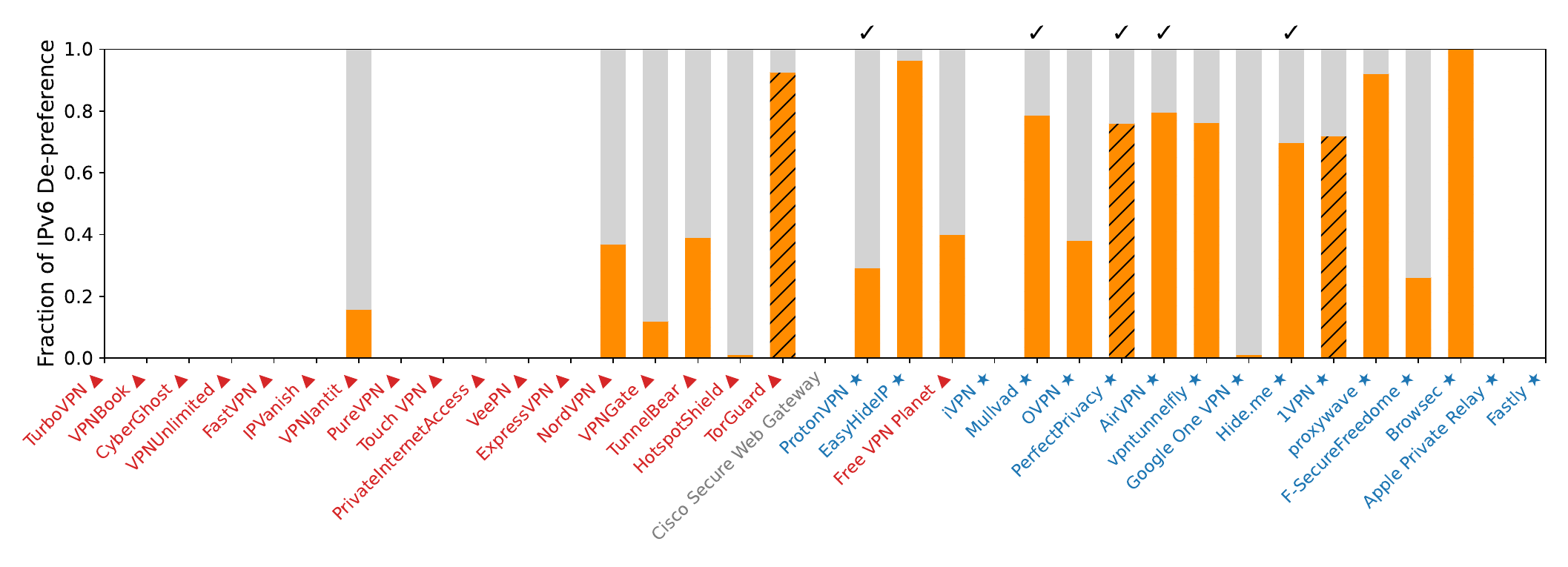}
    \caption{
    Fraction of dual-stack, safe, VPNs sessions to \WIMIA that de-preference IPv6
      for VPNs with at least 20 such sessions per day.
      VPNs marked with a check were tested in \autoref{sec:testbed-validation-of-depreference}
    }
    \label{fig:v6_depreference_pervpn_from_our_data}
\end{figure*}

Surprisingly, we find that \emph{54\% of dual-stack VPN visitors prefer IPv4
  over IPv6}, suggesting that Happy Eyeballs is not working
  for these users.
\autoref{fig:v6_depreference_pervpn_from_our_data} shows the fraction
  of IPv6 depreferencing by VPN provider,
  broken out by VPN.
We report fractions only for VPNs that see
  at least 20 dual-stack, safe sessions per day;
  54\% is the traffic-weighted mean of these fractions.

 % \iffalse
 % \fi
  
By contrast,
  only 13\% of all dual-stack, non-VPNed users who
  prefer IPv4 because the Happy Eyeballs algorithm intentionally prefers
  IPv6~\cite{schinazi_happy_2017}.

\subsection{Protocol Preferences and a Root Cause}
	\label{sec:ipv6_depref_cause}

While on the surface
  Happy Eyeballs
  races IPv4 and IPv6, with a head-start for IPv6,
  HE only comes in to play when equivalent IPv4 and v6 addresses are available.
The Root Cause of VPN depreferencing of IPv6
  is that private addresses are not considered equivalent in IPv4 and IPv6.
  
IP address prioritization is evaluated based on RFC-6724~\cite{rfc6724},
  which classifies and ranks all possible (source, destination) pairs.
(We summarize these rules in \autoref{sec:ipv6_depref_prioritzation_rules_explained}.)
When accessing the Internet without a VPN,
  clients usually use \emph{private} IPv4 and public IPv6 addresses.
RFC 6724 prioritizes public IPv6 over private IPv4, so IPv6 addresses appear earlier in the sorted address list. 
Happy Eyeballs then initiates connection attempts in that order, giving the first address a head start. 
As a result, IPv6 is usually selected, and IPv4 is used only if IPv6 fails or is significantly slower.

When running a VPN, the VPN often assign \emph{private} 
  IPv4 and IPv6 addresses to VPN's tunnel,
  using RFC-1918 space for IPv4 and a ULA address for IPv6.
Here,  with private IPv4 and ULA IPv6 addresses, 
  RFC-6724 prioritization produces a different outcome.
De-prioritization occurs because private IPv4 addresses are treated differently
  from ULA IPv6 addresses---ULA addresses are considered ``link local''
  and their use is discouraged,
  so HE only tries these IPv6 address after a delay.

Thus different treatment of RFC-1918 IPv4 space and ULA IPv6 space
  with standard prioritization rules \emph{systematically de-prioritize
  v6 use for VPN users}.

\subsection{Testbed Validation of De-Preference}
\label{sec:testbed-validation-of-depreference}
To confirm our understanding of VPN de-preferencing of IPv6
  we test 6 commercial VPNs that advertise support of native v6.
We install each vendor's VPN client on an Android device, activate the VPN,
  then inspect the assigned tunnel addresses using the Android Debug Bridge.

We find that one out of the six VPN services (ProtonVPN)
  uses a GUA address on the tunnel interface. ProtonVPN mostly used IPv6 for connection (28\% de-preference).
Although they use GUAs, there is still slightly more depreferencing
  than the 13\% baseline for non-VPN users.

The other five providers we tested
 (Mullvad, AirVPN, hidemeVPN, Perfect Privacy, and Anonine)
  all have very high de-preference fractions: 78\%, 79\%, 69\%, 75\%, and 100\%, respectively.
We see that each of these VPNs assign ULAs,
  and so they frequently de-preference IPv6.
(However, because it is a race, sometimes IPv6 still wins.)  

These results suggest our analysis of how HE and prioritization 
  interact with IPv6 above is a plausible explanation
  for the IPv6 de-prioritization we observe in our data.

\subsection{Options to Avoiding De-Preferencing}
	\label{sec:ipv6_depref_options}

We see three potential solutions to IPv6 de-preferencing:
  (1) VPNs can use GUAs for both sides of their tunnel,
  (2) address prioritization can treat IPv4 and IPv6 private addresses
    similarly,
or (3) we can add a new address class for VPN IPv6 addresses.

While all of these choices are possible,
  for practical reasons we recommend choice (3) and describe it in \autoref{sec:ipv6_depref_solution}.
  
During our discussion of option (2) with members of the IETF who are currently revising RFC 6724, 
they indicated that changing the prioritization of ULAs would be undesirable, as ULAs are defined as non-global addresses.

Although VPNs could use public IPv6 addresses (GUAs),
  distributing GUA addresses to VPN clients would complicate their implementation
and require they treat IPv6 differently than IPv4
  (where private RFC-1918 space is the only option),
  and VPNs like the non-routability of private addresses.

Due to these reasons, it seems unlikely that many VPNs will adopt solution (1).

\subsection{Re-preferencing: a New Address Class}
	\label{sec:ipv6_depref_solution}

We propose to avoid IPv6 de-preferencing in VPNs by
  defining a new, VPN-specific address class,
  a \emph{Tunnel Local Address} or TLA.
These addresses are intended to be used only as VPN tunnel endpoints,
  but should be prioritized the same as IPv4 private addresses,
  thus avoiding de-prioritization while retaining strict separation
  (and non-routability) of VPN-internal addresses.

Currently ULA space is allocated fc00::/7,
  but it uses only fd00::/8.
We suggest TLA could use currently unallocated fc00::/8.
(Alternatively, it could take part of ULA space, like fcf0::/12).

We then suggest updates to RFC-6724 treat TLA space as
  label 1 (same class as GUA), precedence 35 (below GUA, above IPv4).

\textbf{Our prototype:}
We have prototyped TLA
  in Linux-6.15.8 using prefix fc00::/8.
We then modified Linux address selection rules (in \texttt{/etc/gai.conf})
  to follow our proposed changes to RFC-6724.

With this configuration,
   standard browsers will
  consistently select IPv6
  when connecting to v6-enabled Internet destinations,
  confirming this proposed solution allows VPNs to prioritize IPv6
  while maintaining address separation.

\section{Conclusion}
In this work, we examined two problems of commercial VPN services' IPv6 support ---IPv6 traffic leak and IPv6 de-preferencing. 
Using logs from 129k VPN-using daily visitors to \WIMIA and targeted client experiments, we first demonstrated that many IPv4-only VPNs leak tunnel IPv6 traffic to the ISP, exposing default IPv6 address users intend to hide.
Second, for VPNs that advertise IPv6 support, we showed that dual-stack connections overwhelmingly prefer IPv4: 
54\% of Dual-stack VPN sessions use IPv4, around 4 times higher than non-VPN dual-stack users, which shows 13\%,
and controlled tests on Android reveal that five of six providers assign ULA prefixes to the tunnel, which, under RFC6724, are de-prioritized relative to private IPv4, inducing systematic Happy-Eyeballs failure. 
To address this structural bias, we proposed the Translatable Local Address (TLA) space in fc00::/8, and implemented a Linux-based prototype that integrates TLA via gai.conf and standard IPv6 translation, restoring IPv6 preference without kernel changes. 
Our results indicate that VPNs requires improved handling of IPv6.
\label{page:last_body}

\bibliographystyle{ACM-Reference-Format} 

\bibliography{ref}

\appendix

\section{Ethics}
    \label{sec:ethics}

Our analysis uses data from \WIMIA.
This data includes IP addresses associated with specific computers,
  triggered by user requests to the website.
This data is collected as part of regular site operation
  and is consistent with the site's published privacy policy~\cite{WIMIAPrivacy}.

Our study was reviewed by
  USC IRB (\#UP-25-00838).
IRB identified our work as non-human-subjects research
  because although we know requests are associated with individuals,
  we have no way of identifying specific individuals.
As part of our IRB protocol, we agreed not to deanonymize any IP addresses
and not to redistribute any data containing user IP addresses.

\section{Details about Address Prioritization}
	\label{sec:ipv6_depref_details}

As described in \autoref{sec:ipv6_depref_cause},
  IPv6 de-prioritization occurs because
  of an interaction between VPN address assignment
  and address prioritization rules.
Those rules are surprisingly complicated and subtle.

This appendix looks at the browser selection process (\autoref{sec:ipv6_depref_browser_details})
  and the standard rules for how addresses are prioritized (\autoref{sec:ipv6_depref_prioritzation_rules_explained})

\subsection{Browser Implementations to Select Protocols}
\label{sec:ipv6_depref_browser_details}
Browsers and operating system libraries combine
  to look up addresses and select which to use.
\autoref{tab:browser-impl} lists where these rules are given
  and their implementations for three browsers.

\begin{table*}

\begin{tabular}{ll|p{2cm}|p{2.8cm}p{2.8cm}p{2.8cm}|p{2.8cm}}
 & & \multicolumn{5}{c}{\textbf{step}} \\
 & & 0 & 1 & 2 & 3 & 4 \\
 &
   & User req.
   & DNS query
   & Find source addr
   & Sort
   & Race \\
\hline
 & \textbf{specification} 
  & 
  & RFC 8305: HE \cite{schinazi_happy_2017}
  & \multicolumn{2}{|c|} {RFC 6724~\cite{rfc6724}}
  & RFC 8305: HE \\
\hline
\multirow{3}{*}{\rotatebox[origin=c]{90}{\textbf{browser}}}
 & Safari
  & OS
  & \multicolumn{3}{|c|}{\ldots \quad C library (getaddrinfo) \quad \ldots}
  & browser \\
 & Firefox
  & (socket
  & \multicolumn{3}{|c|}{\ldots \quad C library (getaddrinfo) \quad \ldots}
  & browser 
  \\
 & Chrome
  & syscall)
  & \multicolumn{4}{|c}{\ldots \quad browser \quad \ldots}
  \\
\end{tabular}
\caption{Definition of algorithms and location of their implementations for web brwosers.}
	\label{tab:browser-impl}

  \begin{tabular}{cllp{10.0cm}}
    \textbf{Rank} & \textbf{Source} & \textbf{Destination} & \textbf{Reason} \\
    1\ZeroAst &
    IPv6 GUA &
    IPv6 GUA &
    Same label (1) and high precedence (40); most preferred pair. \\

    2 &
    IPv6 ULA &
    IPv6 ULA &
    Same label (13) and precedence (3);
        for internal communication. \\

    3\ZeroAst &
    IPv4 public &
    IPv4 public &
    Same label (4) and precedence 35; valid global IPv4 path. \\

    4 &
    IPv4 private &
    IPv4 private &
    Same label (4) with reachable private addressing; 
        for LANs. \\

    5\ZeroAst &
    IPv4 private &
    IPv4 public &
    Same label (4) and precedence 35, but requires NAT traversal. \\

    6\ZeroAst &
    IPv6 ULA &
    IPv6 GUA &
    Label mismatch (13 vs.\ 1) causes strong penalty. \\

  \end{tabular}
  \caption{
    Address prioritization rules from RFC-6724~\cite{rfc6724}.
    \textbf{Label} groups addresses into the same class (only equality matters).
    \textbf{Precedence} is a numeric preference score (higher is preferred).
  }
  \label{tab:rfc6724-ranking}

\end{table*}

\subsection{RFC-6724 Address Prioritization}	\label{sec:ipv6_depref_prioritzation_rules_explained}

As described in \autoref{sec:ipv6_depref_cause},
  IP address prioritization is the root cause of IPv6 deprioritization
  in VPNs.

IP address prioritization in modern operating systems and browsers follows
  the guidance of RFC-6724~\cite{rfc6724}.
The RFC defines labels and precedences for IP addresses of different times,
  but we summarize those rules to the 7 ranks listed in
  \autoref{tab:rfc6724-ranking}.
Since VPNs route only to global addresses, only ranks
  1, 3, 5, and 6 (marked with *) are relevant to our analysis.

Deprioritization occurs because VPNs using private addresses
  place IPv4 in ranks 5 and IPv6 in rank 6.
With different ranks, only IPv4 is considered for Happy Eyeballs
  and IPv6 is ignored.

% vpn comparison table

\section{Leak Evaluation}

\subsection{Leak Evaluation of All VPN and Prior Studies}
	\label{sec: comparison_table}

In \autoref{sec:related_work_leaks} we discuss prior studies of VPN leaks.
Here we compare VPNs covered in those studies in detail,
  showing the growth of coverage in subsequent studies.

In \autoref{tab:combination-counts} compare coverage by VPN from each study.
Our study sees 123 different VPNs, more than prior studies
  (although we do not see 48 they considered).

More importantly, \autoref{tab:combination-counts}
  highlights that our study shifts the focus
  from binary evaluation (leak or non-leaks) in prior work
  to the fraction of website visitors that show leaks (in our work).
In 35 VPNs (marked with \% in the ``us'' column)
  we are able to evaluate the fraction of visitors that leak,
  and another 100 VPNs (``U'' under ``us'') we see but do not have a large enough sample to report
  a fraction.

\begin{table*}
\resizebox{2.2in}{!}{%
\begin{tabular}{lcccc}
\textbf{Provider} & \textbf{2014} & \textbf{2018} & \textbf{2022} & \textbf{Us (2025)} \\
1ClickVPN & - & - & - & U \\ % (5.71\%) \\
1VPN & - & - & - & 0.03\% \\
4ebur.net & - & - & - & U \\ % (8.84\%) \\
AdGuard & - & - & - & U \\ % (0.00\%) \\
AirVPN & Y & - & N & 0.17\% \\
Algo & - & - & N & - \\
AngelVPN & - & - & - & U \\ % (8.57\%) \\
Anonine & - & - & N & U \\ % (4.51\%) \\
AnonymousVPN & - & - & - & U \\ % (34.00\%) \\
Apple Private Relay & - & - & - & 0.00\% \\
Astrill VPN & Y & - & Y & U \\ % (17.96\%) \\
Atlas VPN & - & - & N & U \\ % (8.11\%) \\
Avast Secureline & - & - & N & U \\ % (6.55\%) \\
Avira Phantom & - & - & N & - \\
Azire VPN & - & - & N & U \\ % (0.53\%) \\
BelkaVPN & - & - & - & U \\ % (100.00\%) \\
Best Proxy Switcher & - & - & - & U \\ % (0.00\%) \\
BestVPN & - & - & N & - \\
Betternet & - & - & N & - \\
BlancVPN & - & - & - & U \\ % (5.50\%) \\
Blaze VPN & - & - & - & U \\ % (0.00\%) \\
BolehVPN & - & - & N & U \\ % (0.00\%) \\
BoxPN & - & - & - & U \\ % (5.30\%) \\
Browsec & - & - & - & 0.01\% \\
BTGuard & - & - & - & U \\ % (15.93\%) \\
Buffered VPN & - & Y & - & - \\
BulletVPN & - & Y & - & U \\ % (10.21\%) \\
Bullguard & - & - & N & - \\
BullVPN & - & - & - & U \\ % (10.33\%) \\
Cactus VPN & - & - & N & U \\ % (12.03\%) \\
CCryptoVPN & - & - & - & U \\ % (0.74\%) \\
Celo & - & - & - & U \\ % (20.00\%) \\
Cisco Secure Web Gateway & - & - & - & 3.89\% \\
Cryptostorm & - & - & N & U \\ % (0.95\%) \\
CyberGhost & - & - & N & 14.31\% \\
EarthVPN & - & - & - & U \\ % (0.00\%) \\
EasyHideIP & - & - & - & 3.35\% \\
elr1c & - & - & - & U \\ % (0.00\%) \\
embracevpn & - & - & - & U \\ % (6.06\%) \\
Encrypt.me & - & - & N & - \\
ExpressVPN & Y & - & N & 6.47\% \\
F-Secure Freedome & - & - & N & 0.01\% \\
FastestVPN & - & - & N & U \\ % (13.70\%) \\
Fastly & - & - & - & 0.00\% \\
FastVPN & - & - & - & 9.53\% \\
FlowVPN & - & - & - & U \\ % (48.24\%) \\
Free VPN & - & - & N & - \\
Free VPN Planet & - & - & - & 1.97\% \\
FreeOpenVPN & - & - & - & U \\ % (14.02\%) \\
freeprovpn & - & - & - & U \\ % (6.50\%) \\
FreeVPN & - & - & - & U \\ % (5.88\%) \\
FrootVPN & - & - & - & U \\ % (4.36\%) \\
GhostPath & - & - & - & U \\ % (16.58\%) \\
Google Fi VPN & - & - & - & U \\ % (1.18\%) \\
Google One VPN & - & - & - & 0.08\% \\
Goose VPN & - & - & N & U \\ % (38.02\%) \\
Hide My Ass & Y & - & N & U \\ % (4.55\%) \\
Hide.me & - & - & N & 0.04\% \\
HideIPVPN & - & Y & N & U \\ % (27.27\%) \\
HideMyIP & - & - & - & U \\ % (2.02\%) \\

\end{tabular}
}
\resizebox{2.2in}{!}{%
\begin{tabular}{lcccc}
\textbf{Provider} & \textbf{2014} & \textbf{2018} & \textbf{2022} & \textbf{Us (2025)} \\
holaproxy & - & - & - & U \\ % (7.36\%) \\
Home \& Away VPN & - & - & - & U \\ % (32.88\%) \\
Hotspot Shield & Y & - & N & 5.54\% \\
IBVPN & - & - & - & U \\ % (0.04\%) \\
IPBurger VPN & - & - & - & U \\ % (27.27\%) \\
IPVanish & Y & - & N & 9.47\% \\
ishaanvpn & - & - & - & U \\ % (0.22\%) \\
Ivacy VPN & - & - & N & U \\ % (6.43\%) \\
IVPN & - & - & N & 1.28\% \\
Jego & - & - & - & U \\ % (0.00\%) \\
K2VPN & - & - & N & - \\
Kaspersky & - & - & N & - \\
KeepSolid & - & - & - & U \\ % (75.00\%) \\
KeepSolid VPN & - & - & N & 11.64\% \\
Le VPN & - & Y & N & U \\ % (22.37\%) \\
LimeVPN & - & - & - & U \\ % (100.00\%) \\
LiquidVPN & - & Y & - & - \\
MaxiVPN & - & - & - & U \\ % (1.96\%) \\
Mozilla VPN & - & - & N & - \\
Mullvad & N & - & N & 0.35\% \\
MyExpatNetwork & - & - & - & U \\ % (29.83\%) \\
myiphider & - & - & - & U \\ % (1.05\%) \\
Namecheap & - & - & N & - \\
Netshade & - & - & - & U \\ % (0.00\%) \\
Njalla & - & - & - & U \\ % (0.00\%) \\
NordLayer & - & - & - & U \\ % (6.80\%) \\
NordVPN & - & - & N & 6.39\% \\
Norton Secure VPN & - & - & Y & - \\
OpenVPN Access Server & - & - & N & - \\
Outline & - & - & N & - \\
OVPN & - & - & N & 0.34\% \\
Panda VPN & - & - & N & - \\
Perfect Privacy & - & - & N & 0.33\% \\
Perimeter81 & - & - & - & U \\ % (25.13\%) \\
personalVPN & - & - & - & U \\ % (13.92\%) \\
PhantomPeer VPN & - & - & - & U \\ % (0.00\%) \\
PIA & N & - & N & 8.95\% \\
Private Tunnel & - & Y & N & - \\
Private VPN & - & Y & N & - \\
PrivateVPN & - & - & - & U \\ % (12.93\%) \\
Proton VPN & - & - & N & 3.40\% \\
proxywave & - & - & - & 0.02\% \\
Psiphon & - & - & N & U \\ % (0.42\%) \\
PureVPN & Y & - & N & 9.10\% \\
RedshieldVPN & - & - & - & U \\ % (3.39\%) \\
Riseup & - & - & N & - \\
SaferVPN & - & - & - & U \\ % (5.86\%) \\
Safum & - & - & - & U \\ % (0.00\%) \\
SandVPN & - & - & - & U \\ % (0.00\%) \\
Seed4.me & - & Y & - & - \\
ShellFire & - & - & - & U \\ % (0.00\%) \\
SmartDNSProxy & - & - & - & U \\ % (10.89\%) \\
Snap & - & - & - & U \\ % (100.00\%) \\
Specdify & - & - & N & - \\
Spotflux & - & - & - & U \\ % (0.00\%) \\
Star VPN & - & - & N & - \\
Steganos & - & - & N & - \\
\end{tabular}
}
\resizebox{2.2in}{!}{%
\begin{tabular}{lcccc}
\textbf{Provider} & \textbf{2014} & \textbf{2018} & \textbf{2022} & \textbf{Us (2025)} \\
StreamVia & - & - & - & U \\ % (0.00\%) \\
Streisand & - & - & N & - \\
StrongVPN & Y & - & N & U \\ % (4.76\%) \\
super\_unlimited & - & - & - & U \\ % (5.88\%) \\
SurfEasy & - & - & Y & - \\
SurfShark & - & - & N & U \\ % (10.86\%) \\
TeklanVPN & - & - & - & U \\ % (100.00\%) \\
Telleport & - & - & - & U \\ % (7.19\%) \\
titantunnelvpn & - & - & - & U \\ % (0.00\%) \\
TorGuard & N & - & N & 5.51\% \\
Touch VPN & - & - & N & 9.09\% \\
Troywell VPN & - & - & - & U \\ % (7.32\%) \\
Trust VPN & - & - & - & U \\ % (3.43\%) \\
Trust.Zone & - & - & N & U \\ % (7.51\%) \\
TunnelBear & Y & - & N & 5.61\% \\
Turbo VPN & - & - & Y & 56.85\% \\
UltraSurf VPN & - & - & - & U \\ % (6.51\%) \\
Unlocator & - & - & - & U \\ % (6.96\%) \\
Unspyable & - & - & N & - \\
Urban VPN Desktop & - & - & N & - \\
uVPN & - & - & - & U \\ % (7.12\%) \\
VanishedVPN & - & - & - & U \\ % (0.00\%) \\
VeePN & - & - & N & 7.01\% \\
VPN Bridge & - & - & - & U \\ % (5.13\%) \\
VPN Hotspot & - & - & N & - \\
VPN Owl & - & - & N & - \\
VPN Plus & - & - & N & - \\
VPN Pro & - & - & N & - \\
VPN Proxy Master & - & - & N & - \\
VPN Super & - & - & N & - \\
VPN.ac & - & - & N & U \\ % (26.85\%) \\
VPN.Asia & - & - & - & U \\ % (0.00\%) \\
VPN.ht & - & Y & - & U \\ % (0.00\%) \\
VPN99 & - & - & - & U \\ % (5.97\%) \\
VPNAccount & - & - & - & U \\ % (0.00\%) \\
VPNArea & - & - & - & U \\ % (0.00\%) \\
VPNBook & - & - & N & 18.44\% \\
VPNGate & - & - & - & 5.93\% \\
VPNhack & - & - & - & U \\ % (7.46\%) \\
VPNJantit & - & - & - & 9.34\% \\
VPNLite & - & - & N & - \\
vpnproxymaster & - & - & - & U \\ % (37.75\%) \\
VpnShop & - & - & - & U \\ % (5.99\%) \\
VPNTunnel & - & - & - & U \\ % (2.27\%) \\
vpntunnelfly & - & - & - & 0.11\% \\
VPNUK & - & - & N & U \\ % (26.42\%) \\
VyprVPN & N & - & N & U \\ % (11.93\%) \\
Windscribe & - & - & N & U \\ % (11.00\%) \\
WiTopia & - & - & - & U \\ % (16.88\%) \\
Working VPN & - & - & - & U \\ % (6.06\%) \\
WorldVPN & - & Y & - & U \\ % (5.90\%) \\
X-VPN & - & - & - & U \\ % (7.82\%) \\
ZenMate & - & - & N & - \\
ZoogVPN & - & Y & N & U \\ % (11.63\%) \\
\end{tabular}
}
\caption{Comparing leaks across all VPNs across our study (last column) and prior studies 
  in 2015~\cite{perta_glance_2015},
%   2016~\cite{ikram_analysis_2016},
  2018~\cite{khan_empirical_2018}, 
  and 2022~\cite{ramesh_vpnalyzer_2022},
  and our study (the rightmost column).
  2015 study tested 5 OSes(PC, mobile), 2018 covered only macOS, 2022 study tested in macOS and Windows. U stands for Undersampled (less than 100 observations per day).
  }
\end{table*}

\begin{table*}[t]
\begin{tabular}{cccc|rr}
\textbf{2014} & \textbf{2018} & \textbf{2022} & \textbf{Us (2025)} & \textbf{Count}  & \textbf{Subtotals} \\
\hline
-- & -- & N & -- & 29 &\\
-- & Y & -- & -- & 3 & \multirow{2}{*}{36} \\
-- & -- & Y & -- & 2 & \\
-- & Y & N & -- & 2 &\\
  \hline
-- & -- & -- & U & 74 & \\
-- & Y & -- & U & 3 & \\
-- & -- & N & U & 16  &\\
-- & Y & N & U & 3 & 100\\
N & -- & N & U & 1 & \\
Y & -- & N & U & 2 & \\
Y & -- & Y & U & 1 & \\
  \hline
-- & -- & -- & \% & 13 &\\
-- & -- & N & \% & 12 &\\
N & -- & N & \% & 3 & 35\\
Y & -- & N & \% & 6 &\\
-- & -- & Y & \% & 1& \\
\hline
13  &  11 & 78 & 135 & 171 & \textbf{171} \\
\end{tabular}
\caption{Combination of outcomes of VPN leak reporets studies
  in 2015~\cite{perta_glance_2015},
  2018~\cite{khan_empirical_2018}, 
  and 2022~\cite{ramesh_vpnalyzer_2022},
  and our study (the rightmost column).}
	\label{tab:combination-counts}
\end{table*}

\subsection{Test Specifications}
\label{sec: test_vpn_specifications}
On MacOS, we tested VPNBook and VPNGate, and in both cases observed IPv6 leakage. 
For VPNBook, the provider distributes multiple OpenVPN and PPTP configuration files, but our tests were limited to one configuration file, \texttt{CA149} OpenVPN bundle. 
For VPNGate, which offers a wide range of servers and protocols, we tested only the top-listed server \texttt{public-vpn-40.opengw.net} using OpenVPN on macOS. 
We also tested TurboVPN on both Android and iOS clients, where we again observed IPv6 leakage.

\label{page:last_page}

\end{document}